# Dual band, low profile and compact tunable frequency selective surface with wide tuning range

Yazdan Rahmani-Shams,[1] Sajad Mohammd-Ali-Nezhad,[1] Ali Nooraei Yeganeh,[1] and Seyed Hassan Sedighy[2,a)]
[1]*Electrical and Electronics Engineering Department, University of Qom, Qom, Iran*
[2]*School of New Technologies, Iran University of Science and Technology, Tehran, Iran*

In this paper, a dual polarized, dual band, low profile, embedded bias network, and compact tunable varactor band pass frequency selective surface (FSS) is designed with a wide tuning range. The proposed FSS is composed of two metallic layers printed on both sides of the substrate, where the top side has two cross strips with different dimensions encircled with high pass grids, whereas the bottom side includes a proper bias network. The loaded varactors between the grids and the cross strips with a designed bias network achieve two independent tunable pass bands. Moreover, an equivalent circuit of the FSS is extracted, which show good agreement with the full wave simulation results. The proposed FSS can operate from 2.28 GHz to 4.66 GHz and from 5.44 GHz to 11.3 GHz by proper tuning of the varactors loaded in the large and small cross strips. The tunable range of the low and high pass bands are about 70% with respect to the center frequency of each pass band. The electrical dimensions of the proposed FSS are about $0.05\lambda \times 0.05\lambda$, where $\lambda$ is the free space wavelength at a lower pass band (2.28 GHz). Moreover, the proposed FSS works properly up to 60° incident angles. *Published by AIP Publishing.* https://doi.org/10.1063/1.5023449

## I. INTRODUCTION

Frequency selective surfaces (FSSs) can be considered as spatial microwave filters with three-dimensional frequency transfer functions made of periodic unit cells, which can be used in absorbers,[1] linear to circular polarization converters,[2] reflectors,[3] spatial filters,[4–7] shielding[8] and radomes.[9] The multi-band FSSs with independent transmission bands are required in many applications such as satellite communication systems. There are different techniques to design multi band FSSs such as genetic algorithms[10,11] fractal elements,[12] multi-element unit cells,[13,14] non-resonant elements,[15] and using complementary structures.[16,17]

In recent years, tunable FSSs have been considered and developed in multifunctional and multi-standard communication systems, multiband reflector antennas, and variable transmission windows in radomes.[18] The tunable FSSs can be categorized into two types: slow tune FSSs and agile tune FSSs. In slow tune FSSs, mechanical changes are usually used for frequency tuning. The mechanically tunable figures and liquid based tunable FSSs are famous members of this group.[19–21] The liquid based tunable FSSs which have a low tuning speed and range are suitable only for special applications. A tunable FSS with a tunable folding angle has been investigated in Ref. 22 which is sensitive to the incident angle and the slow tuning speed.

The electric or magnetic tuning structures (varactors and the ferrite substrate) are usually used in agile tune FSSs.[23–27] Although the ferrite based FSS has a wideband tunable range, their magnetic bias network has some complexity. Moreover, the ferrite elements do not have very fast response which decrease the tuning speed. The varactor based tunable FSSs or electronically tunable FSSs have a faster tuning range rather than other types as well as they are of low cost and have a simple structure and a wide tuning range. In Ref. 23, an electronically tunable band pass FSS has been designed with 65% tuning range and a bulky three dimensional structure which works only in one polarization. Another varactor tunable band pass FSS has been reported in Ref. 24 with a nearly constant bandwidth in the tuning range. However, this structure which is composed of two dielectric layers does not have a wide tuning range. It should be mentioned that FSSs with a single layer substrate have an easier fabrication process, a lower volume and a lower dielectric loss rather than ones with multi-substrate layer FSSs.

To the best of authors' knowledge, the only reported dual band electronically tunable FSS has been proposed in Ref. 16 with small dependence on the pass bands and 55% tuning range. This structure includes two resonators on both sides of the substrate to achieve compact dimensions. However, this FSS response is unstable over the incident angles. In this paper, a new dual pass band frequency selective surface is designed which consists of two metallic layers (FSS unit cells and a bias network) printed on both faces of one thin dielectric substrate. The two groups of varactor diodes are used in these FSS unit cells to tune the pass bands, independently. A proper embedded bias network is proposed which can control the pass bands, independently. The proposed FSS structure operates from 2.28 GHz to 4.66 GHz and from 5.44 GHz to 11.3 GHz band with 70% tuning range and a stable wide response up to 60° incident angles. Moreover, an equivalent circuit model is extracted for the proposed FSS which shows good agreement with the full wave simulation ones. The electrical dimensions of this

[a)]Author to whom correspondence should be addressed: sedighy@iust.ac.ir. Tel. +982173225824.

compact FSS are about $0.05\lambda \times 0.05\lambda$, where $\lambda$ is the free space wavelength at a lower pass band (2.28 GHz).

The paper outline is as follows: the basic design procedure and the unit cell design are discussed in Sec. II. Moreover, the bias network of this structure and the equivalent circuit model are also presented in this section. The simulation and analysis of the proposed FSS is considered in Sec. III. The effect of the bias network and the incident angle in the FSS transmission response is also discussed in this section.

## II. BASIC DESIGN PROCEDURE

### A. Basic idea

The band pass response can be basically designed with two different FSS unit cells, "cross with high pass grid" and "complementary of cross". In the "cross with high pass grid" shown in Fig. 1(a), the cross strips with a band stop response and a grid unit cell with a high pass response are combined to achieve the complex unit cell with the band pass response. In the "complementary of cross" unit cell shown in Fig. 1(b), the cross slots can also be used to achieve the band pass response. These two unit cell types are compared with full wave simulation to achieve the compact band pass FSS. Note that the full wave simulations in this paper are performed by CST microwave studio. The slotted cross results in a band pass resonance at the effective wavelength equal to the cross length marked in Fig. 1(b) by Ls. In the strip cross combined with the grid, the resonance frequency is much lower than the slotted cross where the strip length is almost equal to the quarter of effective length. Notice that this prediction has large variations controlled by the strip width ($w$) and the strip and grid gaps ($g$) that are optimized in the full wave simulation.

The dimensions of these unit cells are similar as $L \times L = 20\,\text{mm} \times 20\,\text{mm}$, while the length and the width of the cross strips (slots) are $Ls = 18.4\,\text{mm}$ and $W = 2.9\,\text{mm}$, respectively. Moreover, the gap considered has $g = 0.3\,\text{mm}$ and $w_g = 0.5\,\text{mm}$ and the copper thickness is 0.018 mm. These unit cells are printed on a RO4350B dielectric substrate with 0.422 mm thickness and $\varepsilon_r = 3.48$. Notice that the manufacturing tolerance of this substrate is about 38 $\mu$m, which does not affect the designed FSS performance. The simulation results of "complementary of cross" and "cross with high pass grid" unit cells are compared in Fig. 1(c). It is obvious that the "cross with high pass grid" has a lower pass band than the "complementary of cross" which means that the "cross with high pass grid" unit cell achieves more compactness. Therefore, this unit cell is chosen as the basic element for the proposed FSS.

The electric field distribution of the "cross with high pass grid" unit cell is shown in Fig. 2 with TM normal incident at the resonant frequency (3.4 GHz) which shows the high electric field at the horizontal cross strip edges with the grid. Therefore, the resonance frequency of the unit cell can be changed by tuning the equivalent capacitance of the gap space between the grid and the strip edges easily. This capacitance can be tuned by varying the strip widths and the space gap. Moreover, it can be tuned by loading the gap with varactor diodes. This fact is used in the proposed tunable FSS structure in Sec. II B.

### B. Unit cell design

To achieve a tunable dual band FSS unit cell, two "cross with high pass grid" unit cells with dimensions can be used as shown in Fig. 3. This structure has two problems: the varactor loading of the high pass unit cell is not achievable due to the high gap distance in the higher band and the bias line network cannot be distributed between the cells easily.

To deal with these problems, a new combination of these two unit cell is proposed as depicted in Fig. 4 in detail. This structure consists of two metallic layers on both sides of the RO4350B dielectric substrate with 0.422 mm thickness. The top layer consists of a periodic arrangement of two crosses with different lengths combined with a high pass circular network in a special manner. The low pass and high pass cross strips are specified by blue and red colors, respectively, where the grids with orange color surround the low pass and high pass strips properly. Also, the surface mounted varactors loaded at the edge of cross trips (high and low pass) are specified by black color. The varactor used in the design is MA46H120 from MACOM Company which is a GaAs constant Gamma Flip-Chip with enough small dimensions to mount at the specified places in the FSS structure. The bias lines networks (green color) are distributed between the unit cells in the bottom layer to bias the varactors through the vias properly. In more detail, each cross strip center (low and high pass) is connected to the bias network through vias. The important property of this bias network is that it almost does not affect the resonant frequencies and pass bands of the proposed structure as discussed in Sec. III in more detail.

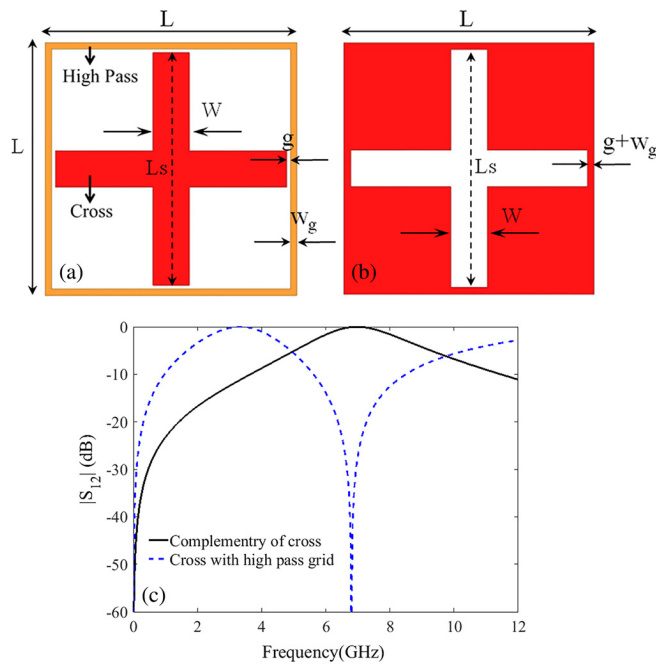

FIG. 1. (a) "Cross with high pass grid" unit cell, (b) "complementary of the cross" unit cell and (c) the unit cell pass band transmission response comparison.

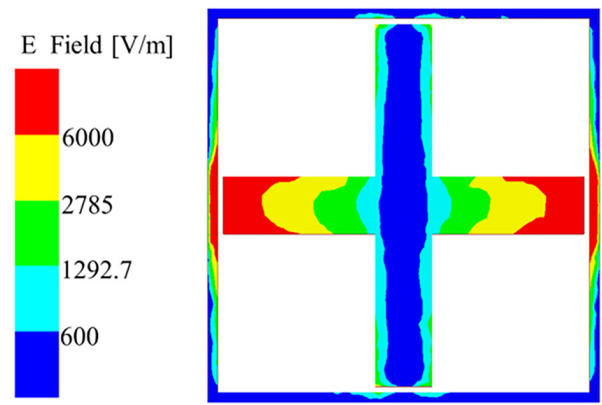

FIG. 2. Electic field distribuition at the resonant frequency (3.4 GHz) with TM polarization incidence.

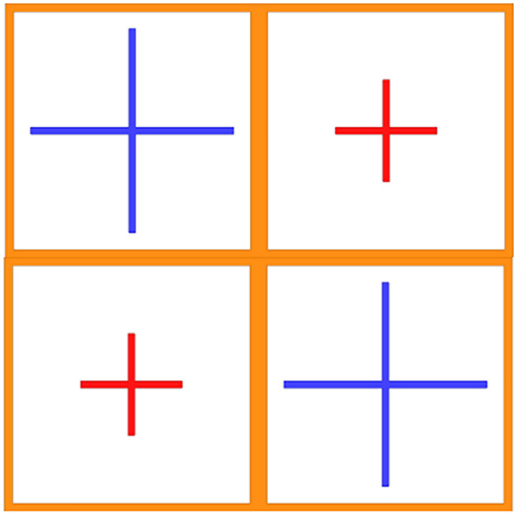

FIG. 3. Dual band cross unit cell composed of two different dimension unit cells and high pass grids.

Now, the main issue is that why the circular grids are chosen instead of the square ones, which are discussed and studied in Sec. II A. For better discussion, the current distributions of these two unit cells (with square grids and with circular grids) are depicted in Fig. 5 at the high pass bands, where the dimensions are similar. At the resonance frequency of small cross, another current is distributed in the high pass square network in addition to the current distributed in small cross strips as shown in the figure. This additional leakage current, which is very higher in the square grid case, changes the effective capacitance between the cross strip edges and the high pass grid, which results in a non-stable transmission response to the oblique incidence wave at a high pass frequency. Therefore, the circular grid structure is chosen in the proposed FSS which has very lower leakage current level as it is clear in the figure, also. The varactors connected to the larger and smaller crosses should be turned on/off independently to control each pass band independent of the other one. Therefore, the surface mounted ac-blocks are used in the bias network as depicted in Fig. 4 and discussed in Sec. II C in more detail.

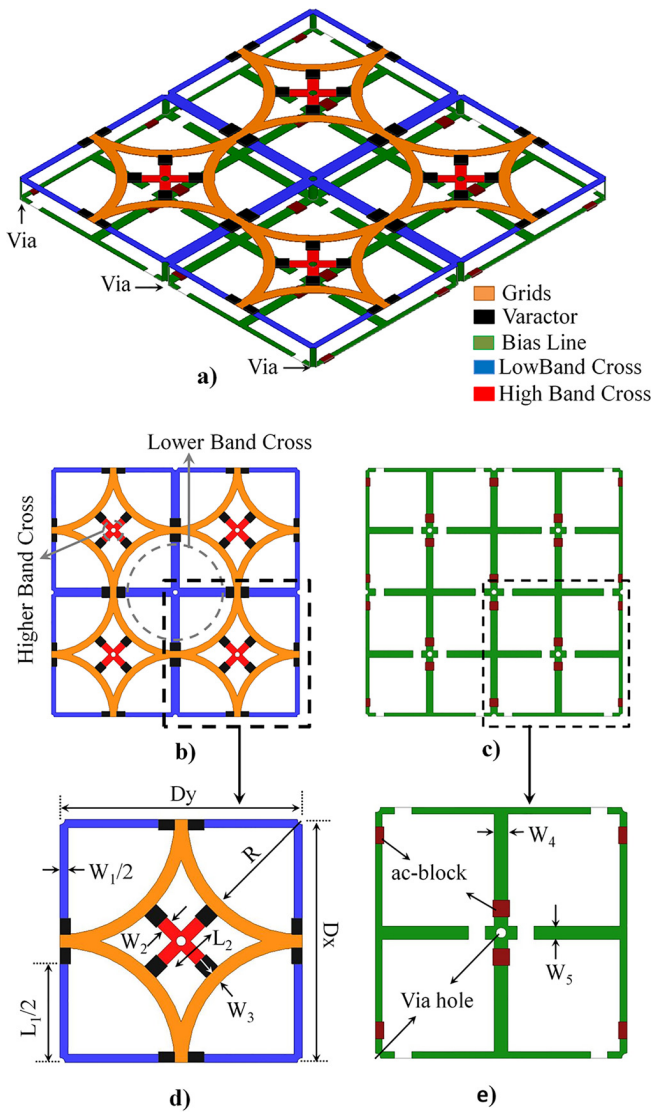

FIG. 4. (a) 3D-view of the four cells. (b) Top view of the FSS four unit cell FSS. (c) Bottom view of the proposed four unit cell FSS. (d) Top view of one unit cell and (e) Bottom view of one unit cell.

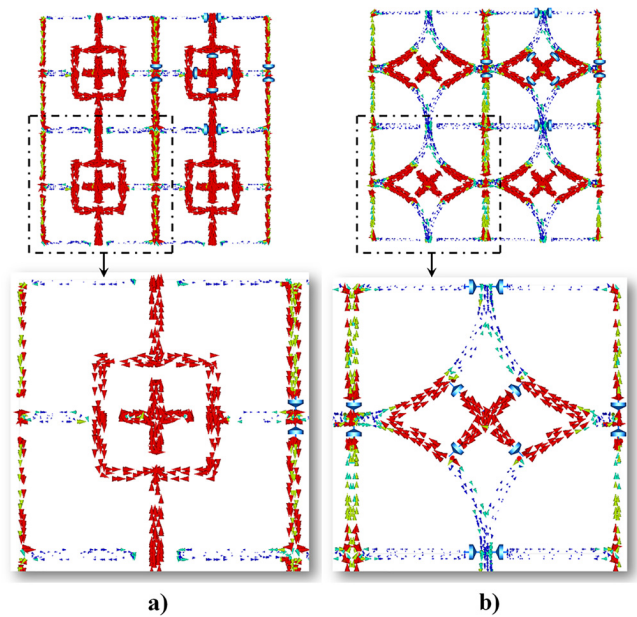

FIG. 5. The current distribution at a high pass frequency in two unit cells with (a) square grids and (b) circular grids.

## C. Bias network

The bias network is composed of two independent DC networks, the first one is designed to control the low pass varactors, while the second one is designed to control the high pass ones. The low pass bias network is controlled by a single DC voltage line printed on the top side of the network which is connected to the low pass strips in the columns through vias to the top layer as depicted in Fig. 6. In more detail, the low pass DC-bias paths start from the top and pass through the ac-blocks. Then, they are transferred to the top layer through the via connected to the center of low pass cross strips. Now, the varactors loaded at the end of these strips are biased and the DC-current moved to the circular grids. The top grids are grounded at one point by via and the ac-block to close the DC bias current paths with RF/DC isolation. Therefore, the low pass bias circuit path is closed without affecting the high pass bias varactors properly. The high pass bias network is designed in a similar manner where the bias control line is placed on the bottom side of the FSS. The DC current passed through the via and turn on the corresponding high pass varactors. This current pass is closed through the grids connected to the ground. Notice that the ac-blocks are used to fracture the bias network and decrease its effect in the FSS behavior. The DC-current for low and high pass bands is depicted in Fig. 6 by black and red arrows, respectively.

## D. Equivalent circuit analysis

An equivalent circuit model of the designed FSS without a bias network is presented in Fig. 7 for normal incidence. The designed FSS has two transmissions zero resulted from the small and large cross strips modeled by two branches, where each branch includes a series inductor with three parallel capacitors. The series equivalent inductance ($L_C$) corresponds to the length of the resonator cross arm, $C_C$ is the capacitance between two adjacent cross arm edges, $C_{cH}$ represents the capacitive effect between the strip edge and the grid, and $C_V$ is the capacitive effect of varactors connected to the crosses. According to the additional current leakage discussed in Fig. 5, there is mutual coupling between the low and high pass cross strips which are modeled by $k_2$. Notice that the footnotes "1" and "2" in the circuit model stand for the low and high pass cross strips, respectively. $L_H$ and $C_H$ also model the high pass behavior of the grids. Notice that the colors used in this model correspond to the color used in Fig. 4 where orange is used for grids, blue for the low pass band, red for the high pass band and black for the varactors. Since the high pass grids affect the small cross resonance frequency which is reduced by changing the grid shape into circular, the inductor mutual coupling is modeled by $k_1$ between $L_{H2}$ and $L_{C2}$ in the circuit model.

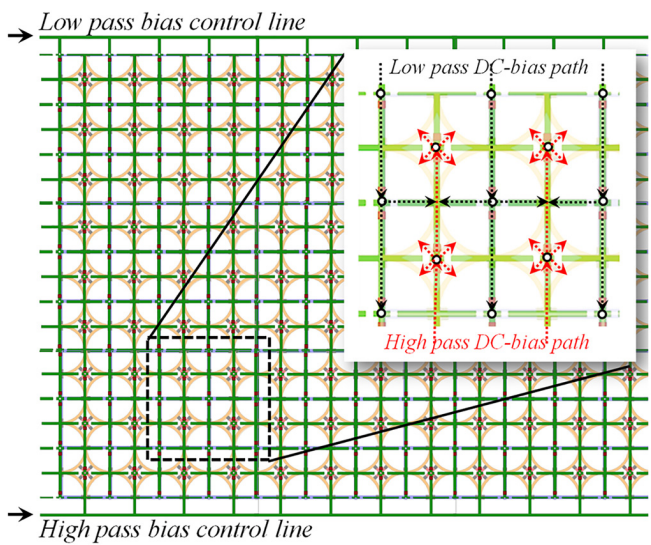

FIG. 6. The bias network details: the DC-current for low and high pass bands is depicted with black and red arrows, respectively.

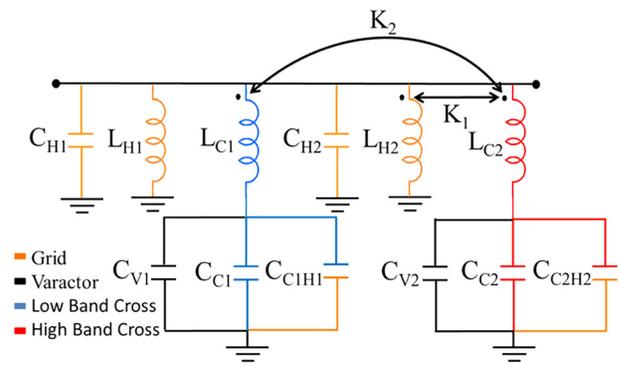

FIG. 7. Equivalent circuit model of the proposed FSS structure.

## III. SIMULATION AND ANALYSIS

The proposed FSS is designed to work at two independent bands printed on a Rogers 4350B with 0.422 mm thickness and a copper thickness of 0.018 mm. The geometric dimensions of the structure are: $D_x = D_y = 7.2$ mm, $a = b = 0.6$ mm, $W_1 = W_2 = W_3 = 0.2$ mm, $R = 3.5$ mm, $W_4 = W_5 = 0.2$ mm, $L_1 = 6$ mm, $L_2 = 1.78$ mm. The full wave simulations are carried out in CST Microwave Studio by using the Floquet boundary condition to replicate the infinite planar array of unit cells.

For scrutinization of the bias network effect on the proposed FSS transmission, the simulation results of the FSS

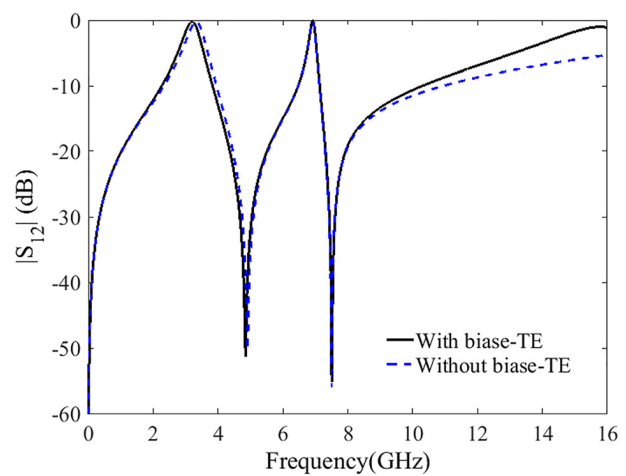

FIG. 8. Transmission responses of the designed tunable FSS under normal incidence ($\theta = 0°$) with and without a bias network.

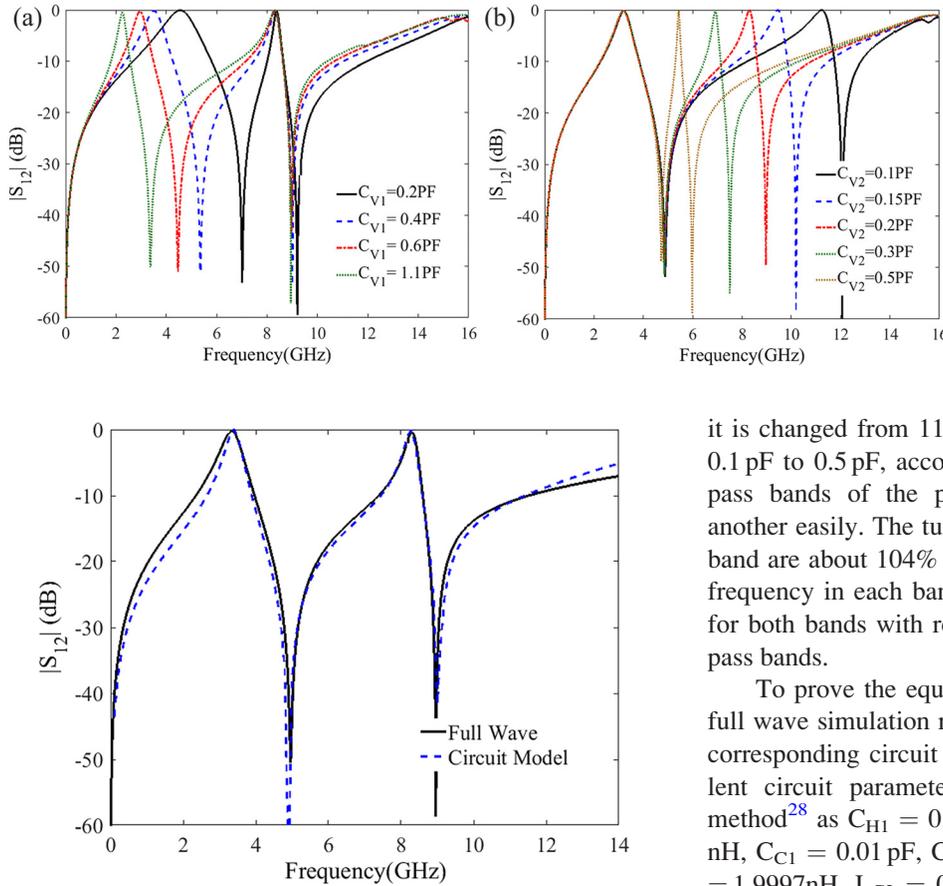

FIG. 9. (a). Transmission responses of the designed tunable FSS under normal incidence ($\theta = 0°$) for different values of varactors: (a) $C_{V1}$ is varied from 0.2 pF to 1.1 pF, where $C_{V2}$ is 0.2 pF and (b) $CV_2$ is varied from 0.1 PF to 0.5PF, where $C_{V1}$ is 0.5 pF.

FIG. 10. The transmission response of the equivalent circuit model and the full wave simulation of the designed FSS.

with and without the bias network are compared in Fig. 8 for TE polarization incident with varactor loading of the cross strips. The results clearly prove the negligible effect of the bias network on the FSS transmission response, which is suitable for the design of the structure. This is the reason why the bias network is not included in the modeled circuits discussed in Sec. II D.

The pass band of the proposed FSS is independently tunable by using suitable varactors. Figure 9(a) depicts the low pass band variations from 4.66 GHz to 2.28 GHz by changing the varactor capacitance from 0.2 pF to 1.1 pF, where the high pass band cross varactors are set to 0.2 pF for the normal TE incident wave. The high pass band can be easily tuned by changing the varactor as shown in Fig. 9(b) where it is changed from 11.3 GHz to 5.44 GHz by variation from 0.1 pF to 0.5 pF, accordingly. Therefore, one can tune each pass bands of the proposed FSS independent from one another easily. The tunable ranges for the low and high pass band are about 104% and 107.7% with respect to the lowest frequency in each band. Moreover, this value is about 70% for both bands with respect to the center frequency of each pass bands.

To prove the equivalent circuit proposed in Sec. II, the full wave simulation results of the proposed unit cell and its corresponding circuit are compared in Fig. 10. The equivalent circuit parameters are derived with a curve fitting method[28] as $C_{H1} = 0.0052$ pF, $L_{H1} = 1.48$ nH, $L_{C1} = 1.024$ nH, $C_{C1} = 0.01$ pF, $C_{C1H1} = 0.001$ pF, $C_{H2} = 0.19$ pF, $L_{H2} = 1.9997$ nH, $L_{C2} = 0.73$ nH, $C_{C2} = 0.056$ pF, $K_2 = 0.165$, $K_1 = 0.49$, $C_{C2H2} = 0.24$ pF, $C_{V1} = 1$ pF, $C_{V2} = 0.4$ pF, and are performed using Advanced Designed System (ADS) software. The incident wave is considered normal to the FSS with TE polarization, also. As it is clear, there is good agreement between the full wave and circuit model results which prove the proposed model, properly.

The sensitivity of the proposed FSS transmission response to the oblique incidences is evaluated for both TE and TM polarized waves. The simulation results are depicted in Fig. 11 for $\theta = 0°$, 30°, and 60° at both TE and TM polarizations with two different representative varactor loading cases, $C_{V1} = 0.5$ pF, $C_{V2} = 0.15$ pF and $C_{V1} = 0.2$ pF, $C_{V2} = 0.1$ pF, where the bias network is also considered in the simulations. It is observed that the designed FSS depicts stable pass band transmission responses for the incident angles up to 60° at both polarizations which prove the capability of the proposed FSS rather than the references.

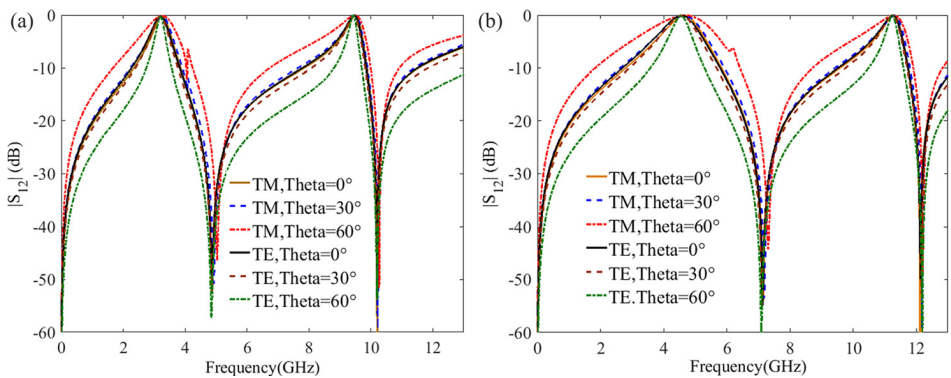

FIG. 11. Transmission responses of the proposed tunable FSS for oblique incidences and TE and TM polarizations, at $\theta = 0°$, 30°, and 60°. (a) $C_{V1} = 0.5$ pF, $C_{V2} = 0.15$ pF (b) $C_{V1} = 0.2$ pF, $C_{V2} = 0.1$ pF.

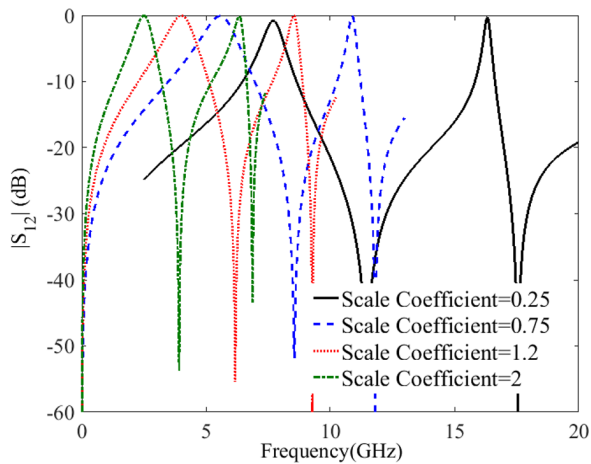

FIG. 12. The transmission response of the proposed unit cell with different scale coefficients.

As an important specification, the operation bands of the proposed FSS can be easily tuned by scaling the proposed unit cell dimensions. For example, the transmission responses of the proposed unit cell with different scale coefficients are depicted in Fig. 12 for a TE polarized incident wave, where $C_{v1} = 0.4\,\text{pF}$ and $C_{v2} = 0.2\,\text{pF}$. It is clear that the required FSS unit cell with desired operation bands can be easily designed by scaling.

## IV. CONCLUSION

A varactor tunable dual polarized, low profile, dual band and compact FSS has been designed, where its pass bands can be tuned independent of each other. The varactor bias network was designed to provide independent band tuning capability without any undesired effect on the FSS operation. An equivalent circuit has been extracted for the proposed FSS, which showed good agreement with the full wave simulation. The tunable range of low and high pass bands is about 70% with respect to the center frequency of each pass band. Moreover, this FSS depicted stable pass band transmission responses for the incident angles up to 60° at both polarizations.